\documentclass[sigconf]{acmart}
\usepackage{multirow}%
\usepackage{amsfonts}%
\usepackage{amsthm}%
\usepackage[title]{appendix}%
\usepackage{textcomp}%
\usepackage{algorithm}%
\usepackage{listings}%
\usepackage{algorithmic}
\usepackage{url}
\usepackage{multicol}%%%
\usepackage{subcaption}
\usepackage{color}
\usepackage{colortbl}
\begin{document}

\title[Social Biases in Knowledge Representations of Wikidata]{Social Biases in Knowledge Representations of Wikidata separates Global North from Global South}


\author{Paramita Das}
\email{paramita.das@iitkgp.ac.in}
\affiliation{%
  \institution{Dept. of CSE, IIT Kharagpur}
  \country{India}}

\author{Sai Keerthana Karnam}
\email{karnamsaikeerthana@iitkgp.ac.in}
\affiliation{%
  \institution{Dept. of CSE, IIT Kharagpur}
  \country{India}}

\author{Aditya Bharat Soni}
\email{adityabs@cs.cmu.edu}
\affiliation{%
  \institution{Carnegie Mellon University}
  \country{USA}}

\author{Animesh Mukherjee}
\email{animeshm@cse.iitkgp.ac.in}
\affiliation{%
  \institution{Dept. of CSE, IIT Kharagpur}
  \country{India}}

\renewcommand{\shortauthors}{P. Das, et al.}

\begin{abstract}
  Knowledge Graphs have become increasingly popular due to their wide usage in various downstream applications, including information retrieval, chatbot development, language model construction, and many others. Link prediction (LP) is a crucial downstream task for knowledge graphs, as it helps to address the problem of the incompleteness of the knowledge graphs. However, previous research has shown that knowledge graphs, often created in a (semi) automatic manner, are not free from social biases. These biases can have harmful effects on downstream applications, especially by leading to unfair behavior toward minority groups. To understand this issue in detail, we develop a framework -- \textbf{AuditLP} -- deploying fairness metrics to identify biased outcomes in LP, specifically how occupations are classified as either male or female-dominated based on gender as a sensitive attribute. We have experimented with the sensitive attribute of age and observed that occupations are categorized as young-biased, old-biased, and age-neutral. We conduct our experiments on a large number of knowledge triples that belong to 21 different geographies extracted from the open-sourced knowledge graph, Wikidata. Our study shows that the variance in the biased outcomes across geographies neatly mirrors the socio-economic and cultural division of the world, resulting in a transparent partition of the \textit{Global North} from the \textit{Global South}.
\end{abstract}




\iffalse

\fi

%\ccsdesc[500]{Do Not Use This Code~Generate the Correct Terms for Your Paper}
%\ccsdesc[300]{Do Not Use This Code~Generate the Correct Terms for Your Paper}
%\ccsdesc{Do Not Use This Code~Generate the Correct Terms for Your Paper}
%\ccsdesc[100]{Do Not Use This Code~Generate the Correct Terms for Your Paper}



\maketitle

\section{Introduction}
The rapid growth of online content has led to the widespread adoption of knowledge graphs (KG) by both industry giants~\cite{yahya2021semantic} and academic researchers. KGs are being utilized in various applications, e.g., language model development~\cite{logan2019baracks,agarwal2021knowledge}, question answering~\cite{saxena2020improving,huang2019knowledge}, personalized recommendation~\cite{xu2021fair} and others. A knowledge graph is a structured database that contains information about real-world knowledge in the form of nodes representing real-world entities and directed edges with labeled semantic relations connecting the nodes. The open source knowledge graphs, such as DBpedia, Wikidata, %\footnote{\url{https://github.com/dbpedia}}, Wikidata
%\footnote{\url{https://www.wikidata.org/wiki/Wikidata:Main_Page}}, YAGO
%\footnote{\url{https://yago-knowledge.org/}} 
etc., have been grown to the web-scale level, consisting of millions of entities and billions of facts. However, it is widely acknowledged that even the largest and most comprehensive KGs have limitations in terms of incompleteness, containing a fraction of the vast amount of real-world knowledge. To address this issue of incompleteness, link prediction (LP) techniques utilize the existing facts in KG to make predictions about new facts. \\
\noindent\textbf{Bias in KG:} Unfortunately, while high-quality structured content is a plus, a wide range of societal and human biases are inherent to KGs in many ways -- be in the form of sampling strategy or the judgmental view. A pertinent question in the research community exists regarding the source of biases in the knowledge graph. Historical facts contain demographic biases that lead to an imbalance of knowledge triples between different social groups. Further, a notable study by~\cite{demartini2019implicit} explored how paid crowdsourcing can be investigated to understand the implicit biases in the beliefs of individuals can influence curating the knowledge graphs. This work highlights the underlying fact that knowledge graphs inherently encompass the pre-existing biases and assumptions of their editors, who shape, modify, and remove the data within the knowledge graph. %Another work by~\cite{janowicz2018debiasing} has pointed out the types of biases that become apparent within knowledge graphs. 
In addition to the inherent biases as above, there can be amplifications or alterations of social biases during the generation of the dense representation of the entities and relations in the knowledge graph, aka KG embedding~\cite{kraft2022lifecycle}. 
%Here it is important to recognize the interplay between the biases that are inherently present in KGs and those that emerge during the embedding learning process. 
In~\cite{das2023diversity}, the authors showed that the data bias existing in the knowledge graph from its inception can undergo further alterations during the embedding learning phases, specifically due to the characteristics of the learning algorithms, referred to as algorithm bias. With the proliferation in the usage of KG-based models in widely deployed systems such as search engines, chatbot applications, etc., social biases and their harmful prejudices are propagated to these applications. For instance, if a KG predominantly features data about men in certain occupations, it may reinforce stereotypes, suggesting those roles are meant only for men. Thus, it is crucial to systematically analyze these biases, particularly in tasks related to knowledge completion, such as link prediction~\cite{kraft2022lifecycle} and algorithmic decision making~\cite{lahoti2019ifair,bartley2021auditing}. Researchers have proposed various approaches to measure bias by leveraging the structural information within knowledge graphs~\cite{bourli2020bias, fisher2020measuring, du2022understanding}. Other studies treat link prediction as a classification task, further examining biases by analyzing the classifier's outcomes using different fairness metrics~\cite{keidar2021towards, schwertmann2023model}. In this context, it is worth mentioning that KGs face challenges when balancing bias versus fairness. Through a systematic analysis, authors in \cite{rossi2021knowledge} identified three types of data biases that can affect link prediction datasets and models. This experiment aimed to estimate the extent to which biased data can affect the accuracy of link prediction models in the completion of KGs. Another research has investigated the issue of structural imbalances in KGs, particularly in the context of KG completion~\cite{shomer2023toward}. Our work measures biases at the micro level for each individual human entity using the fairness metrics of \textit{equal opportunity} and \textit{equalized odds}. In addition, it highlights patterns of biases across aggregate social groups (i.e., geographies), bridging a gap that has been largely overlooked in the existing literature.\\
%researchers~\cite{arduini2020adversarial,bose2019compositional,fisher2020debiasing} tried to invent various methodologies, such as adversarial learning to mitigate biases from knowledge graphs which in turn can help to design a bias-free automated system for different downstream tasks.}
%In addition to detecting and addressing biases, researchers have also looked into the impact of biases on tasks such as link prediction used for KG completion. . \\
\noindent\textbf{Present work:} We believe that our work is the first one in the literature that considers a large corpus extracted from a knowledge graph (Wikidata), comprising knowledge triples from 21 geographies spanning multiple continents and investigates the biases that creep in while learning representations using various algorithms. Our study focuses on link prediction as a downstream task to measure the biases in Wikidata. We pose two key research questions to align our contributions in seeking answers--
\begin{itemize}
    \item \textbf{RQ1:} How do social biases (i.e., gender: male/female and age: young/old in our work) impact predictive outcomes related to an observable (occupation or profession in our case) in the link prediction task given a knowledge graph?
    \item \textbf{RQ2:}  How do biases vary by varying the geo-social data used in downstream link prediction, and whether the patterns of biases so revealed point to some universal characterization of the geographies?
\end{itemize}
We extract triples from 21 different geographic locations around the globe, including Arabia, India, Israel, Japan, South Korea, Turkey, Russia, Australia, New Zealand, Egypt, Nigeria, South Africa, France, Germany, Spain, United Kingdom, Argentina, Brazil, Canada, Mexico and the United States of America. To investigate how social biases affect the fair link prediction of occupations, we propose a noble framework named \textbf{AuditLP} to measure the biased predictions given the sensitive attributes of human entities. Our framework is built upon popular knowledge graph embedding learning (KGE) algorithms -- TransE~\cite{bordes2013translating}, DistMult~\cite{yang2015embedding}, CompGCN~\cite{shang2019end}, and GeKC~\cite{loconte2024turn} which generate knowledge graph embeddings as the features to be used in the prediction. The framework is designed to hide relationship links (between human entities and occupation names) from the training instances while learning the embeddings by the KGE algorithms. Next, the learned embeddings, aka features, are passed to a classifier to predict the links. Finally, the predictions are analyzed in light of the notion of fairness. Analyzing the prediction outcomes by \textit{AuditLP}, we came across the following key findings --
\begin{itemize}
    \item The classifier outcomes are unfair regarding sensitive attributes for a given set of professions.
    \item Our study also reveals that the choice of data, i.e., geographies, significantly impacts the variance of biases. Specifically, the social biases present in different geographies manifest into a clear partition of the world into two distinct regions: the Global North and the Global South, characterized by their different geo-social and economic attributes\footnote{\url{https://en.wikipedia.org/wiki/Global_North_and_Global_South}}. Surprisingly, this result is true for all the algorithms we used in \textit{AuditLP} -- TransE, DistMuslt, CompGCN, and GeKC-- even though their inner workings are quite different. Such a result indicates that this observation has a universal underpinning.
\end{itemize}

\noindent All the codes and data of the work are made available\footnote{\url{https://github.com/paramita08/Wikidata_AuditLP}} for reproducible research.

\section{Related Work} \label{sec:related_work}

%We will now briefly discuss the key highlights of knowledge graph embedding, link prediction, and biases in knowledge graphs and their impact on digital applications upon which our research is built.\\ 
Knowledge graph embedding (KGE) techniques have become very popular due to their extensive use in many downstream applications. Numerous KGE algorithms have been developed in recent years, among which \textit{TransE}~\cite{bordes2013translating}, and its variants, \textit{RESCAL}~\cite{nickel2011three}, \textit{HolE}~\cite{nickel2016holographic}, \textit{CrossE}~\cite{zhang2019interaction}, \textit{ComplEx}~\cite{trouillon2016complex}, \textit{ConvKB}~\cite{nguyen2017novel}, \textit{ConvE}~\cite{dettmers2018convolutional}, \textit{HypER}~\cite{balavzevic2019hypernetwork} etc. are being used hugely. On the other hand, various methods~\cite{coscia2021multilayer,kumar2020link,zhang2020learning} have been proposed to perform link prediction, relying on different embedding representation techniques~\cite{wang2021survey,rossi2021knowledge}. 
Despite the importance of link prediction, biases can occur due to various factors such as incomplete data, imbalanced training data, or the presence of implicit biases in the training data or the link prediction algorithm itself~\cite{wang2022degree}. Such biases can lead to inaccurate or unfair predictions, particularly in applications such as recommendation systems or decision-making algorithms~\cite{chen2023bias}. Social biases, such as gender bias, can lead to various challenging problems, particularly in large peer-production platforms like Wikipedia~\cite{beytia2022visual,cabrera2018gender,wagner2015s}. Similar biases have also been observed in the Wikidata in terms of gender~\cite{zhang2021quantifying}, race, and country of citizenship~\cite{shaik2021analyzing} as well as in various class-levels~\cite{ramadizsa2023novel}. As a follow-up, researchers~\cite{arduini2020adversarial,bose2019compositional,fisher2020debiasing} have tried to invent various methodologies, such as adversarial learning, to mitigate these biases from knowledge graphs. While most of the above works are US-centric, in our work, we consider as many as 21 different geographies and conduct a systematic audit of how biases creep in while learning representations of knowledge triples from Wikidata using various algorithms, how these biases affect the link prediction task and show universal patterns in the way they manifest across geographies dividing them into the Global North and the Global South.  

\section{Dataset description} \label{sec:dataset}
Wikidata is an open-source knowledge base that serves as a repository for structured data used in various Wikimedia projects like Wikipedia and Wikivoyage~\cite{vrandevcic2023wikidata}. Much like other Knowledge Graphs (KGs), Wikidata organizes information into triples, which consist of a subject item, a property, and an object. The objects in these triples can be either entities or literals, such as quantities or strings. Subject and object items are identified by URIs that start with 'Q' (e.g., Q937 for Albert Einstein), while properties are represented by URIs prefixed with 'P' (e.g., P27 for country of citizenship). We extract a specific set of triples from Wikidata based on certain criteria and conduct experiments using this carefully curated dataset.

%Wikidata serves as a freely available and open-source knowledge base, acting as a repository for structured data for various Wikimedia projects like Wikipedia and Wikivoyage. Similar to other Knowledge Graphs (KGs), information in Wikidata is stored in the form of triples, consisting of a subject item, a property, and an object. Objects can be entities or literals, such as quantities or strings. Subject and object items are represented by URIs beginning with 'Q' (e.g., Q5284 for \textit{Bill Gates}), while properties are denoted by URIs prefixed with 'P' (e.g., P19 for \textit{place of birth}). We extract a specific set of triples from Wikidata based on specific criteria and perform experiments on this curated dataset.

\begin{figure}
    \centering
    \includegraphics[height= 5cm, width=0.5\textwidth]{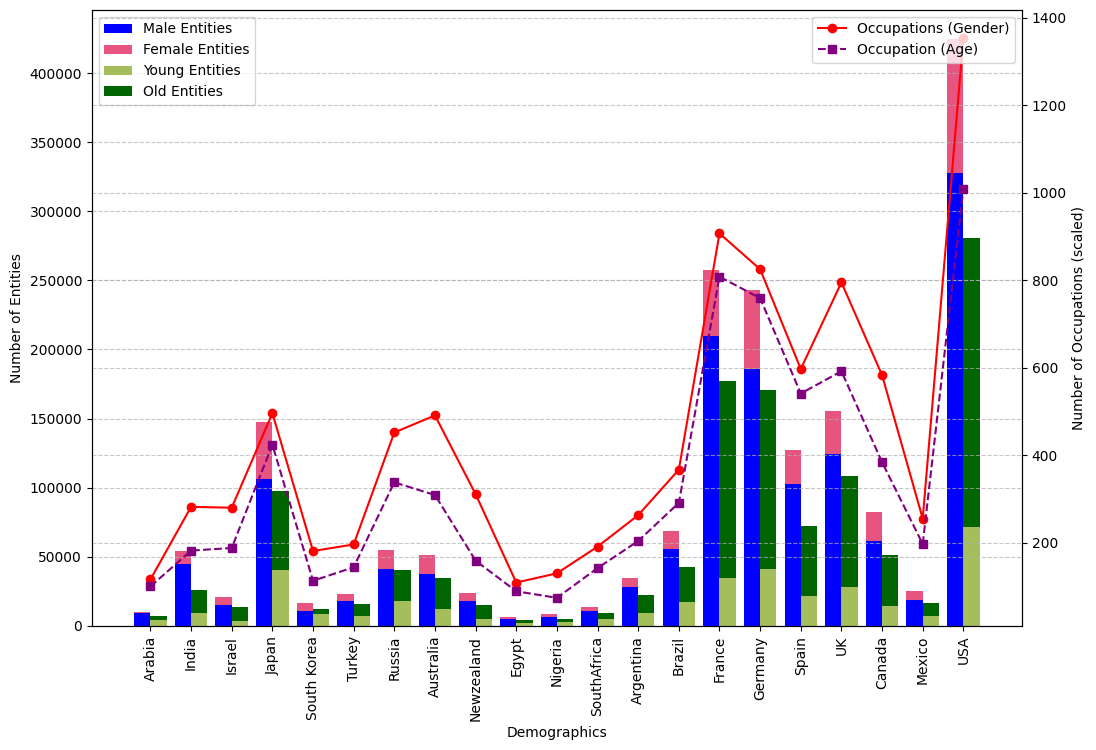}
    \caption{Figure showing the count of human entities (i.e., male/female, young/old, and corresponding number of occupations per geography in our dataset.}
    \label{fig:stat_demos}
\end{figure}

\noindent\textbf{Collection of geography dataset}: 
In our research, we make use of the latest Wikidata dump\footnote{\url{https://dumps.wikimedia.org/wikidatawiki/entities/latest-all.json.bz2}}, which is stored in JSON format, compressed in the bz2 format, taking up approximately 70 GB when compressed. Initially, we process this dump through the widely-used KGTK\footnote{\url{https://kgtk.readthedocs.io/en/latest/}} library to streamline data handling. This step results in the creation of three distinct files: a node file ($\sim9.5$ GB), an edge file ($\sim189.5$ GB), and a qualifiers file ($\sim60.4$ GB). KGTK is a Python library commonly employed to simplify the manipulation of knowledge graphs. Within the node file, one can find information like English labels, descriptions, and aliases linked to both Qnodes and Pnodes.
In contrast, the edge file contains the triplets found within the knowledge graph, along with additional details about the tail entity, such as its language and entity type etc. Subsequently, we undertake a filtration process to exclude triplets that lack a Wikidata identifier (QID) as the head entity or a Wikidata property identifier (PID) as the relation. This filtration results in a dataset comprising an impressive 1.32 billion triplets, encompassing 93 million entities (QIDs) and 8,763 relations (PIDs). Our primary aim is to collect the triples associated with 21 geographic attributes spanning various continents, which we have specifically chosen for our research. These geographies include Arabia, Australia, Argentina, Brazil, Canada, Egypt, France, Germany, India, Israel, Japan, Mexico, New Zealand, Nigeria, Russia, Spain, South Africa, South Korea, Turkey, the United Kingdom, and the United States of America. To identify human entities within a specific geography, we compute the overlap between entities that are both human and associated with that particular geography. To establish this connection, we extract the head entities from the collected triplets that have the relation $P31$ (``instance of'') and the tail entity $Q5$ (``human''). Moreover, the head entities of these triplets must be linked to the corresponding country using the relation $P27$ (``country of citizenship'') and the tail entity representing the QID of that specific country. By doing so, we gather the human entities belonging to each geography. Additionally, we consider all outgoing edges from these human entities and collect all the triplets where the head entity belongs to the set of human entities we obtain earlier. This enables us to collect occupation information for all the entities and to create geography-specific knowledge graphs. For the Arabia dataset, we specifically consider the outgoing edges from human entities belonging to any country within the Arabian Peninsula. 
Now, for every human entity, we extract two sensitive attributes -- gender and date of birth from the Wikidata triples. In the case of gender, we restrict the experiment to the binary division of gender -- male and female -- as only an extremely few human entities of non-binary gender exist in Wikidata. Next, we compute the age of a human entity as per the date of the experiment (i.e., January 2024) and divide them in every geography into two age groups -- `young' and `old.' The persons whose ages are in the range of 19--45 years are identified as young. On the other hand, people, having age in the range 60--90 years are labeled as old. We keep a gap of 15 years between the two ranges so that the distinction between the young and old would be very prominent. A brief statistics of humans -- male/female and young/old and the corresponding number of professions are shown in Figure~\ref{fig:stat_demos}. 
The figure clearly illustrates that there is an imbalance in the proportion of male and female entities as well as young and old entities in each geography. This difference is the result of sampling bias, aka data bias, that mimics the inequalities found in the actual world. Further, the distribution of professions in America or Europe is in stark contrast to those from Asia or Africa. This difference highlights how Western culture dominates other cultures on collaborative platforms like Wikidata. It is important to note that we are careful while choosing the training data for our experiment to maintain the diversity present in the geographic dataset of Wikidata. Our dataset is carefully designed to accomplish the downstream task of link prediction and measure biases in the task. The uniqueness of the large dataset lies in the two factors-- \textbf{(1)} categorization of knowledge triples based on geographies and \textbf{(2)} availability of sensitive information, such as age and gender, about human entities.

\section{Bias and fairness} \label{sec:background}
In a KG, facts are represented as labeled triplets in the form of $<h, r, t>$, where $h$ and $t$ stand for the head and tail entities, respectively, and $r$ represents the relationship between them. KG embedding (KGE) algorithms are used to learn compact representations of entities and relations in a low-dimensional embedding space. In a downstream link prediction (LP) task, the embeddings of the head entity and the relation are combined to identify the most likely tail entity. This is done by training an objective function specific to a KGE model that maps the triples to a scalar score, maximizing the likelihood of correctly predicting the triples. To perform a detailed analysis of biases in the link prediction task, we measure biases across different groups of KGE models and further diversify the experiment by examining Wikidata triples across various geographical divisions. \\
\noindent\textbf{Definition of bias:} Understanding of bias in link prediction is grounded in the concept of representational harm, specifically referring to the differences in system performance for different social groups. For instance, in the case of gender bias in predicting occupations, if a link prediction model is less accurate in predicting the occupations of women compared to men, it would be considered to exhibit harmful behavior. This is because deploying such a biased model in downstream applications could reduce its utility for women compared to men, potentially reinforcing societal inequalities in the predictive performance of present-day AI systems, including the large language models.

\noindent\textbf{Notion of fairness}: The notion of fairness refers to the concept of treating individuals and groups equitably in different aspects, including algorithms, predictive outcomes, etc., in general by the whole automation system. In the context of link prediction, fairness denotes that the predictions should not systematically discriminate against particular individuals or groups based on certain sensitive attributes such as gender, race, religion, age, etc. We replicate this notion of fairness in our experiment by assessing the impact of sensitive attributes of a human entity, such as gender or age, on the accuracy of predictions for target properties, like occupation. Let us assume, in a classification setup for predicting whether a human has a specific occupation or not, the following notations stand for--
\begin{enumerate}
    \item \textit{G}: Protected or sensitive attribute for which (non)discrimination should be established. In our experiment, it is -- (a) gender: male and female (binary division), (b) age: young and old.
    \item \textit{Y}: The actual class (1 or 0 in our case) labels as per the dataset, i.e., whether a particular human entity has the occupation $p$ (label = 1) or not (label = 0). 
    \item \textit{$\overline Y$}: Predicted outcome, i.e., class labels for an individual obtained from the classifier. 
\end{enumerate}
Now, the predictive outcomes of the classifier considered for identifying fair classification are mentioned below. Here, $m$ and $f$ stand for male and female, respectively. On the other hand, young and old people are denoted by $yn$ and $ol$, respectively.

\noindent \textbf{True positive rate (in case of gender)}: 
\begin{equation*}
\scriptsize
    %\resizebox{0.7\hsize}{!}{%
    TPR_m = P(\overline Y = 1 | Y = 1 , G: gender)
   % }
\end{equation*}

\begin{equation*}
\scriptsize
    %\resizebox{0.7\hsize}{!}{%
    TPR_f = P(\overline Y = 1 | Y = 1 , G: gender) 
   % }
\end{equation*}

\noindent \textbf{True positive rate (in case of age)}: 
\begin{equation*}
\scriptsize
   % \resizebox{0.7\hsize}{!}{%
    TPR_{yn} = P(\overline Y = 1 | Y = 1 , G: age)
  %  }
\end{equation*}

\begin{equation*}
\scriptsize
   % \resizebox{0.7\hsize}{!}{%
    TPR_{ol} = P(\overline Y = 1 | Y = 1 , G: age)
  %  }
\end{equation*}

\noindent \textbf{False positive rate (in case of gender)}: 
\begin{equation*}
\scriptsize
    %\resizebox{0.7\hsize}{!}{%
    FPR_m = P(\overline Y = 1 | Y = 0 , G: gender)
    %}
\end{equation*}

\begin{equation*}
\scriptsize
    %\resizebox{0.7\hsize}{!}{%
    FPR_f = P(\overline Y = 1 | Y = 0 , G: gender)
    %}
\end{equation*}

\noindent \textbf{False positive rate (in case of age)}: 
\begin{equation*}
\scriptsize
    %\resizebox{0.7\hsize}{!}{%
    FPR_{yn} = P(\overline Y = 1 | Y = 0 , G: age)
    %}
\end{equation*}

\begin{equation*}
\scriptsize
    %\resizebox{0.7\hsize}{!}{%
    FPR_{ol} = P(\overline Y = 1 | Y = 0 , G: age)
    %}
\end{equation*}
    
\noindent We followed two state-of-the-art fairness metrics-- \textit{Equal Opportunity} and \textit{Equalised Odds} that measure the fair prediction of a classification task, i.e., link prediction in our experiment. 

\begin{itemize}
    \item \textbf{Equal opportunity}: It demands the positive outcome to be independent of the protected attribute, $G$, conditional on Y being an actual positive. In other words, the True Positive Rate (TPR) is to be the same for each element of the protected attribute, e.g., male and female or young and old in our classification setup.
    \begin{equation*}
    \scriptsize
        TPR_{m} = TPR_{f}
    \end{equation*}
    \begin{equation*}
    \scriptsize
        TPR_{y} = TPR_{o}
    \end{equation*}
    \item  \textbf{Equalized odds}: It defines the positive outcome to be independent of the protected attribute G, conditional on the actual Y. In terms of classification outcomes, the True Positive Rate (TPR) and False Positive Rate (FPR) are to be the same for each element of the protected attribute, e.g., male and female or young and old in our classification setting.
    \begin{equation*}
        TPR_{m} = TPR_{f} \hspace{1mm} and \hspace{1mm} FPR_{m} = FPR_{f}
    \end{equation*}

    \begin{equation*}
        TPR_{y} = TPR_{o} \hspace{1mm} and \hspace{1mm} FPR_{y} = FPR_{o}
    \end{equation*}
\end{itemize}

\noindent Based on the above definitions, we label an occupation as either male-biased/female-biased/gender-neutral or young-biased/old-biased/age-neutral, which is introduced in the subsequent sections.

\noindent\textbf{KGE models}: 
Using our proposed framework, AuditLP, we demonstrate social biases in link prediction. To achieve this, we select four widely used embedding learning algorithms -- TransE, DistMult, CompGCN, and GeKC to generate knowledge graph embeddings for the triples in our dataset. The rationale behind choosing these models is that each of them represents a popular genre of embedding learning, and our objective is to investigate the performance of the algorithms of each genre. Our hypothesis is that the other algorithms from each of these genres would behave similarly to the representative ones we chose. According to the standard categorization of KGE models~\cite{wang2017knowledge}, TransE~\cite{bordes2013translating} is a translational approach, DistMult~\cite{yang2015embedding} distance-based semantic matching approach, and CompGCN~\cite{vashishth2019composition} is convolution neural network based approach. GeKC~\cite{loconte2024turn} is the latest addition to the KGE learning paradigm, which uses generative KGE circuits to enhance the efficiency and reliability of triple predictions for the missing link prediction task. Our study aims to demonstrate how models from each of these genres perform on our curated dataset. %Our hypothesis is that models within each genre would yield closely similar results.}

\section{Methodology} \label{sec:methodology}
To investigate how social biases encoded in KGs are further perpetuated to downstream tasks, we proposed our framework named \textit{AuditLP} for predicting the links in a given geographic dataset. We posit link prediction as the classification task in which the classifier is supposed to predict whether a true triple ($<human\_entity$, $have\_occupation$, $occupation\_name>$) has the occupation or not.
The framework comprises four main stages, and the steps are further detailed in the following sections. The framework finds biases across the sensitive attribute-- gender and its division of gender-biased occupations-- male-biased, female-biased, and gender-neutral. An exactly similar approach is followed when measuring biased outcomes in terms of sensitive attribute age, in which the occupations are grouped into three categories-- young-biased, old-biased, and age-neutral.
\begin{figure*}[h]
    \centering
    \includegraphics[height=5cm,width=0.8\textwidth]{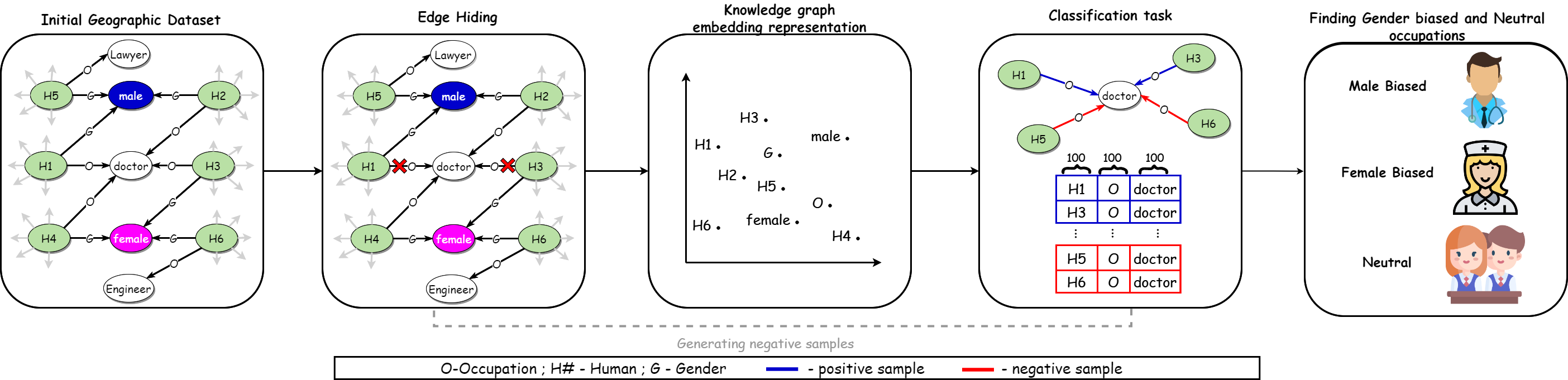}
    \caption{Schematic for our experiments showing different steps in case of sensitive attribute gender-- edge hiding, embedding generation, and classification of occupations.}
    \label{fig:pipeline}
\end{figure*}

\noindent\textbf{Hiding edges}: Each geography present in our dataset has occupations with at least two male and two female human entities, which are considered in our experiment. Since our goal is to predict the occupation of a human entity, the occupation information should not be encoded in the human entity embedding while generating it in the training phase. Therefore, in this step, we hide (i.e., remove) the occupation information of the human entities whose occupations have to be predicted in the classification phase. Let us assume that there exists $x$ number of human entities having occupation $p$ in a geography $d$, with male and female entities as $m$ and $f$, respectively. The triple \textit{$<$ $H_{1}$, occupation, $O_{1}$ $>$} indicates that the human entity $H_{1}$ has the occupation $O_{1}$. We randomly select 50\% of the human entities from $x$ while maintaining the male-female ratio and remove their occupation (triples similar to \textit{$<$ $H_{1}$, occupation, $O_{1}$ $>$}) to get the filtered dataset. The occupation triples removed from the initial graph are treated as positive samples for classification. Further, an equal number of negative samples are generated from the human entities in the original graph that do not have the occupation $p$.\\
\noindent\textbf{Generation of knowledge graph embeddings}: We train the filtered geographic by the embedding learning models-- TransE, DistMult, CompGCN, and GeKC one at a time. For the learning algorithms (except GeKC), we set the dimensions of the graph embedding to 100 and 200 for GeKC. The models are trained for a varying number of epochs depending on the size of the dataset using a fixed negative sample size of 10, with multiclass negative log-likelihood loss as the cost function and SGD as the optimizer. We utilize the python library AmpliGraph\footnote{\url{https://docs.ampligraph.org/en/1.4.0/}} for implementing and generating embedding by TransE and Pykeen\footnote{\url{https://pykeen.readthedocs.io/en/stable/}} in case of DistMult and CompGCN. For GeKC, we utilized the public repository\footnote{\url{https://github.com/april-tools/gekcs}} made by the authors to train the geographies. We evaluated the embeddings in terms of standard metrics like MRR and Hits@$n$ ($n$ is set to 5, 10, and 20) and achieved results that are comparable to benchmark results reported in the literature~\cite{ali2021bringing}. \\
\noindent\textbf{Predicting occupation of human entities}: After generating embeddings of humans, relations, such as `has\_occupation', and occupation entities, our goal is to predict the occupations of human entities. Given a triple of the form \textit{$<$ $H_{1}$, has\_occupation, $O_{1}$ $>$}, the classifier answers whether the human entity $H_{1}$ has the occupation $O_{1}$ or not. In other words, the link prediction classifier takes as input a concatenation of vectors representing -- (gendered) head entity vector $\odot$ relation entity (i.e., `has\_occupation') vector $\odot$ tail entity (i.e., occupation name) vector and predicts a 0 (no relation exists) or 1 (relation exists), based on a set of training instances. For training purposes, positive triples are the triples that have occupation $p$ in the initial dataset, while negative triples are selected from the human entities that do not have occupation $p$. The dimension of the triples is kept at 300, which is obtained by concatenating the embeddings of the human, relation, and occupation entities, each having 100 dimensions. To accomplish the classification, we use an MLP classifier, and an 80:20 split was used between the training and test sets. The classification result, i.e., the true positive rate and the false positive rate, are considered for further analysis. For each embedding algorithm, we perform classification for each of the geographies individually. To measure the classification results based on gender divisions -- male and female, we calculate $TPR$ and $FPR$ for both male and female entities individually, as defined in the previous section.\\

\noindent\textbf{Finding gender-biased occupations}: Now, we attempt to categorize the occupations as gender-dominated or gender-neutral based on the $TPR$ and $FPR$ values as predicted by the classifier. Utilizing the two metrics of fairness-- \textit{Equal Opportunity} and \textit{Equalized Odds} (as discussed in the section~\ref{sec:background}), the occupations in every geography are identified to be in one of three categories -- male-biased, female-biased, and gender-neutral.

\noindent An occupation $o$ is identified as male-biased if it satisfies any one of the two definitions --
\begin{equation} 
%\scriptsize 
TPR_{m}(o) > TPR_{f}(o) \end{equation} 
\begin{equation} 
%\scriptsize
    TPR_{m}(o) = TPR_{f}(o),  FPR_{m}(o) > FPR_{f}(o)
\end{equation}

\noindent Similarly, a female-biased occupation follows any of the two definitions below --
\begin{equation}
%\scriptsize 
    TPR_{f}(o) > TPR_{m}(o) 
\end{equation} 
\begin{equation}
%\scriptsize
    TPR_{m}(o) = TPR_{f}(o),  FPR_{f}(o) > FPR_{m}(o)
\end{equation}

\noindent Naturally, an occupation $o$ is said to be gender-neutral if it satisfies the following constraint. 
\begin{equation} 
%\scriptsize
    TPR_{m}(o) = TPR_{f}(o),  FPR_{f}(o) = FPR_{m}(o)
\end{equation}

\noindent We observe that sometimes the $TPR$ and $FPR$ values are nearly equal for male and female entities for certain occupations in the test dataset, and those are ignored from being categorized as male/female-biased occupations. Henceforth, we consider an occupation as \textit{male-biased} if it satisfies any of the following.

\begin{equation}
%\scriptsize
      TPR_m(o) - TPR_f(o) \ge t_{1} 
\end{equation}
\begin{equation}%\scriptsize
      TPR_m(o) = TPR_f(o) , FPR_m(o) - FPR_f(o) \ge t_{2} 
\end{equation}

\noindent and \textit{female-biased} if any of the following holds true.
\begin{equation}
%\scriptsize
      TPR_f(o) - TPR_m(o) \ge t_{1} 
\end{equation}
\begin{equation}
%\scriptsize
      TPR_m(o) = TPR_f(o), FPR_f(o) - FPR_m(o) \ge t_{2} 
\end{equation}

\noindent For \textit{gender-neutral} occupations, we considered the following.
\begin{equation}
%\scriptsize
    \resizebox{0.99\hsize}{!}{
    $\left|TPR_m(o)-TPR_f(o)\right| < 0.01$ \hspace{0.5em}  
    $\left|FPR_m(o)-FPR_f(o)\right| < 0.01$ 
    }
\end{equation}

\noindent To set the threshold $t_{1}$, we take the differences of $TPR_{m}$ and $TPR_{f}$ for all the occupations and compute $\mu-\sigma$ of these differences. Similarly, we compute $t_2$ based on the $FPR$ differences. Such methods of eliminating outlier values have been abundant in the literature~\cite{reimann20051}. If both $t_{1}$ and $t_2$ are as low as 0.01, we denote such cases as gender-neutral occupations for all geographies alike. Thus, following this pipeline, we have three lists of occupations prepared for every geography --
(i) male-biased, (ii) female-biased, and (iii) gender-neutral.
In the case of sensitive attribute age, we set $\mu + \sigma$ of the $TPR$ and $FPR$ differences as the corresponding thresholds $t_1$ and $t_2$ to cut off all the outlier cases. An occupation is assumed as age-neutral if the absolute differences of $TPR_{yn}$ and $TPR_{ol}$ or $FPR_{yn}$ and $FPR_{ol}$ is lesser than 0.01. In the case of sensitive attribute age, we divide the lists of occupations into three categories of biased occupations-- (i) young-biased, (ii) old-biased, and (iii) age-neutral.

\section{Result} \label{sec:result}

\subsection{Fairness in classification} 

We run the classifier on each of the 21 geographies individually and iterated the process for TransE, DistMult, CompGCN, and GeKC. We obtain accuracy (aka Acc), F1-score, AUC score, $TPR_{m}$, $TPR_{f}$, $FPR_{m}$, $FPR_{f}$ for each occupation, and average the results over all of the occupations and all geographies. We report this average ($\mu$) along with the standard deviation ($\sigma$) for each metric in the case of the training accomplished by TransE, DistMult, CompGCN, and GeKC in Table~\ref{tab:classification_performance} (first seven rows). Similarly, for age, we compute accuracy (aka Acc), F1-score, AUC score, $TPR_{yn}$, $TPR_{yn}$, $FPR_{ol}$, $FPR_{ol}$ for each occupation and the average value over all the occupations and geographies and corresponding standard deviation are tabulated in the last seven rows of Table~\ref{tab:classification_performance}. Overall results show that at an aggregate level, i.e., considering all the geographies together, the average values of $TPR$ and $FPR$ feature slight differences when the males and females are compared. This is true for the age attribute as well. Therefore, it is clear that the prediction outcome is not fully fair and equal across the social groups. However, the picture changes when we observe the occupations at the individual geography level. We utilize equations (6) through (10) to create a list of occupations categorized as either male-biased, female-biased, or gender-neutral for each of the 21 geographies. We have come across various interesting observations where occupations are biased differently in different geographies. For instance, in the USA, occupations like real estate brokerage are male-biased, while occupations like prostitution and pornographic actors are female-biased. We observe that countries like France, Germany, and Spain have the occupations of linguist, engineer, doctor, professor, and economist as male-biased, and the female-biased professions include sports like volleyball, alpine skier, blogger, etc. In contrast, we note that likewise USA, in Germany too, pornographic actor is identified as a female-biased occupation. Further, a few occupations, such as film actors and artists, manifest as gender-neutral in many countries, such as Argentina, Brazil, South Africa, Canada, Mexico, etc. A similar trend is observed in the case of the sensitive attribute age. Laborious occupations, such as tennis player, boxer, war photographer, police officer, etc., are categorized as young-biased across almost all the geographies -- Arabia, South Africa, Nigeria, Canada, Spain, UK, USA, etc. On the other hand, sports commentators, computer scientists, film actors, vocalists, and other less labor-intensive occupations are labeled as old-biased occupations. Further, some age-neutral occupations in different countries include basketball players in the USA, researchers in France, songwriters in India, dub actors in Israel, poets in Mexico, etc. Although geographies with a larger number of triples, such as the USA, Germany, and France, contain occupations in almost all categories in contrast to smaller geographies such as Arabia, Israel, etc., overall, there is a wide variance in the biased occupations for both gender and age across all these geographies. Sample biased occupations are provided\footnote{\url{https://tinyurl.com/3df9hr4z}} for further examination. Occupations categorized (i.e., male/female, young/old) by the fairness metrics give a very detailed picture of the biased professions across geographies, and it is not easy to see a universal pattern of social biases. Therefore, we next outline an approach to automatically extract such patterns.

\begin{table}[h]
%\scriptsize
    \centering
    \begin{tabular}{|p{1.2cm}|p{0.5cm}|p{0.5cm}|p{0.5cm}|p{0.5cm}|p{0.5cm}|p{0.5cm}|p{0.5cm}|p{0.5cm}|}
     \hline
    \multirow{2}{*}{Metric} & \multicolumn{2}{c|}{TransE} & \multicolumn{2}{c|}{DistMult} & \multicolumn{2}{c|}{CompGCN} & \multicolumn{2}{c|}{GeKC} \\
    \cline{2-9}
    & $\mu$ & $\sigma$ & $\mu$ & $\sigma$ & $\mu$ & $\sigma$ & $\mu$ & $\sigma$\\
    \hline
    $TPR_{m}$ & 0.90 & 0.09 & 0.84 & 0.09 & 0.85 & 0.04 &  0.90 & 0.09 \\
    \hline
     $TPR_{f}$ & 0.86 & 0.10 & 0.83 & 0.08 & 0.85 & 0.05 & 0.81 & 0.09 \\
    \hline
     $FPR_{m}$ & 0.15 & 0.11 & 0.17 & 0.14 & 0.13 & 0.03 & 0.07  & 0.03 \\
    \hline
    $FPR_{f}$ & 0.16 & 0.10 & 0.20 & 0.13 & 0.13 & 0.04 & 0.07 & 0.04 \\
    \hline
    $Acc_{gender}$ & 86.50 & 8.43 & 82.66 & 10.99 & 81.59 & 18.70 & 92.97 & 1.24\\
    \hline
    $F1_{gender}$ & 0.86 & 0.8 & 0.80 & 0.12 & 0.85 & 0.03 & 0.89 & 0.04 \\
    \hline
    $AUC_{gender}$ & 0.93 & 0.8 & 0.86 & 0.14 & 0.94 & 0.02 & 0.92 & 0.05 \\
    \hline 
    $TPR_{y}$ & 0.87  & 0.11 & 0.85 & 0.14& 0.82 & 0.05 & 0.92 & 0.04\\
    \hline
    $TPR_{o}$ & 0.85 & 0.10 & 0.83 & 0.16 & 0.83 & 0.04 & 0.88 & 0.03\\
    \hline
    $FPR_{y}$ & 0.15  & 0.14 & 0.18 & 0.16 & 0.16 & 0.04 & 0.06 & 0.03\\
    \hline
    $FPR_{o}$ & 0.19  & 0.13 & 0.22 & 0.18 & 0.14 & 0.04 & 0.04 & 0.01 \\
    \hline
    $Acc_{age}$ & 84.32 & 0.10 & 92.18 & 0.12 & 84.03 & 3.79 & 93.01 & 0.03 \\
    \hline
    $F1_{age}$ & 0.84 & 0.09 & 0.92 & 0.13 & 0.84 & 0.04 & 0.90 & 0.03  \\
    \hline
    $AUC_{age}$ & 0.91 & 0.10 & 0.97 & 0.13 & 0.92 & 0.03 & 0.94 & 0.04 \\
    \hline
    \end{tabular}
    \caption{Table showing the mean and standard deviation of different metrics averaged over all geographies for TransE, DistMult, CompGCN, and GeKC. Here, the first \textbf{seven} rows tabulate metrics computed for gender, and the last \textbf{seven} rows are for age.}
    \label{tab:classification_performance}
\end{table}

\subsection{Social biases at macro level}
To discover universal patterns, we characterize each geography based on a five-dimensional vector representation. Each entry in this vector corresponds respectively to each of the conditions described in equations (6)-(10). For a given occupation if it satisfies one of the conditions then the corresponding vector entry will be 1, others are zero. For a given geography, the individual vector representation for each of the occupations belonging to that geography is added to obtain the geography-level vector. Next, we cluster this vector space using a standard hierarchical spectral clustering that partitions the 21 geographies in the following way. Similar to any clustering algorithm, the number of clusters $k$ is a parameter and is adjusted using the standard elbow method. This whole process is separately executed for all three learning algorithms TransE, DistMult, CompGCN, and GeKC, and the clusters obtained are listed in Table~\ref{tab:clusters_list}. Remarkably, the clusters are broadly consistent with the popular socio-economic partitioning of the world -- the \textit{Global North} and \textit{Global South}~\cite{odeh2010comparative}. This observation is true for all the algorithms TransE, DistMult, CompGCN, and GeKC despite the differences in their inner workings. Typically, countries that are economically developed are part of the Global North, while those that are in the developing phase are considered the Global South. For example, for gender, in the case of TransE, the first two clusters, aka Cluster 1 and Cluster 2 (henceforth GN-1 and GN-2), are dominated by the countries from the Global North while Cluster 3 is from the Global South (henceforth GS). Likewise, gender, in the case of age too, the same division -- Global North and Global South emerges from the clustering for all the KGE algorithms. The clusters for both gender and age are noted in Table~\ref{tab:clusters_list}; the first 3 rows denote clusters obtained in case of gender, and the last 2 rows are for age.\\
\noindent\textbf{Quantitative evidence}: To corroborate that our clusters indeed correspond to the global economic divide we further compute a set of country-level attributes, such as distance from American culture~\cite{muthukrishna2020beyond}, GDP per capita (in USD)\footnote{\url{https://www.imf.org/en/Publications/WEO/weo-database/2023/April}}, Gini coefficient~\cite{de2007income}, Human Development Index~\cite{anand1994human}, Gender Gap\footnote{\url{https://www.weforum.org/reports/}}, and Individualism~\cite{navcinovic2019role}. We report the average values of these predictors for the clusters in Table~\ref{tab:country_predictors}. The reported values show that, indeed, GN-1 and GN-2 have significantly different values for the above attributes compared to the GS and reveal significant differences between GN-1, GN-2, and GS, particularly in terms of the meanings conveyed by these attributes. For instance, the attribute American Cultural Distance shows that GN-1 has the lowest values, indicating that the geographies within GN-1 share very similar cultural characteristics with that of the USA. In contrast, GS exhibits much higher values, reflecting a greater cultural difference from the USA. We use average cosine similarity of the features of the countries in a cluster to measure intra-cluster similarities. The higher this similarity, the more close the countries within a cluster are.\\ 
\begin{table}[h]
    \centering
    \scriptsize
    \begin{tabular}{|p{0.3cm}|p{1.5cm}|p{1.5cm}|p{1.5cm}|p{1.5cm}|}
    \hline
     & \textbf{TransE} & \textbf{DistMult} & \textbf{CompGCN} & \textbf{GeKC} \\
     \hline
      $C_{1}$: & Argentina, Australia, Canada, France, Germany, Israel, Japan, Russia, Spain, UK, USA  &  Australia, Canada, France, Germany, Japan, Russia, Spain, UK, USA & Australia, Spain, France, Canada, UK, Japan, Germany, Russia, USA, Brazil & Australia, Argentina, Canada, France, Germany, Israel, Japan, UK, USA \\
      \hline
      $C_{2}$: & Mexico, New Zealand, South Korea & Argentina, Brazil, Israel, Mexico, New Zealand, South Korea &  Arabia, South Africa, South Korea, New Zealand, Argentina, Egypt, India, Nigeria, Israel, Mexico, Turkey & Arabia, Brazil, Egypt, Russia, India, Mexico, Nigeria, New Zealand, Spain, South Africa, South Korea, Turkey\\
      \hline
      $C_{3}$: & Arabia, Brazil, Egypt, India, Nigeria, South Africa, Turkey & Arabia, Egypt, India, Nigeria, South Africa, Turkey &  &  \\
      \hline
      $C_{1}$: & Argentina, Canada, UK, Germany, Japan, Australia, USA, Russia, Brazil, Spain, France  &  Canada, Brazil, Mexico, Russia, USA, UK, France, Israel, Germany, Japan, Spain & Spain, Brazil, France, Australia, Russia, Japan, UK, Germany, Canada, USA & Australia, Argentina, Brazil, Canada, France, Germany, India, Japan, New Zealand, Spain, UK, USA\\
      \hline
      $C_{2}$: & Arabia, South Korea, Israel, Nigeria, Turkey, Mexico, India, Egypt, South Africa, New Zealand & New Zealand, South Africa, Australia, Nigeria, India, Arabia, Egypt, Turkey, Argentina, South Korea & Mexico, Argentina, New Zealand, Arabia, Turkey, Nigeria, South Africa, South Korea, India, Israel, Egypt & Arabia, Egypt, Israel, Mexico, Nigeria, Russia, South Korea, South Africa, Turkey \\
      \hline
    \end{tabular}
    \caption{Table showing different clusters obtained by the clustering of the features. The upper block (i.e., first 3 rows) and the lower block (i.e., last 3 rows) represent the clusters generated in the case of sensitive attributes of gender and age, respectively.}
    \label{tab:clusters_list}
\end{table}
\begin{table*} [ht]
    \centering
    %\scriptsize
  \begin{tabular}{|p{2.2cm}||p{1.0cm}|p{1.0cm}|p{1.0cm}|| p{1.0cm}|p{1.0cm}|p{1.0cm}||p{1.0cm}|p{1.0cm}||p{1.0cm}|p{1.0cm}|}  \hline
    \multirow{2}{*}{Attributes} & \multicolumn{3}{c|}{TransE} & \multicolumn{3}{c|}{DistMult} & \multicolumn{2}{c|}{CompGCN}  & \multicolumn{2}{c|}{GeKC}  \\
    \cline{2-11}
    & GN-1 & GN-2 & GS & GN-1 & GN-2 & GS & GN & GS & GN & GS\\
    \hline
American Cultural Distance ($\downarrow$) & 0.06 & 0.07 & 0.12 &  0.06 & 0.07 &   0.13 & 0.05 & 0.08 & 0.06 & 0.10\\\hline
GDP per capita (in USD) ($\uparrow$) & 44,827.82 & 30,606 & 11,807 & 47,125 & 28,276 & 12,298 & 43,298 & 21,326 & 49,915.89 & 18,194 \\\hline
Gini coefficient ($\downarrow$) & 35.81 & 37.03 & 40.3 & 34.77 & 40.15 & 38.86 & 36.19 & 38.65 & 35.79 & 38.75 \\
\hline
Human Development Index ($\uparrow$) &  0.91 & 0.85 & 0.71 & 0.88 & 0.74 & 0.70 & 0.89 & 0.78 & 0.92 & 0.78\\\hline
Gender Gap Index ($\uparrow$) & 0.75 & 0.74 & 0.67 & 0.75 & 0.73 & 0.67 & 0.74 & 0.69 & 0.75 &  0.71\\\hline
Individualism ($\uparrow$) & 65.82 & 42.33 & 43.28 & 69.33 & 44.16 & 44.16 & 66.2 & 44.72 & 70.44  & 43.34 \\\hline
Intra-cluster similarities ($\uparrow$) & 0.35 & 0.99 & 0.87 & 0.59 & 0.78 & 0.65 & 0.65 & 0.77 & 0.89 & 0.73 \\
\hline
%Inter-cluster distances & (GN2): -0.63, (GS): -0.60 & (GN1): -0.63 ,(GS): 0.93  & (GN1): -0.60 , (GN2): 0.93 &  (GN2): -0.51, (GS):-0.55  & (GN1): -0.51 ,(GS): 0.16  & (GN1): -0.55 , (GN2): 0.16 \\
\end{tabular}
\caption{Different country-level attributes showing social, economic, and cultural differences and intra-cluster similarities computed for the geographies grouped by clusters for sensitive attribute gender}
\label{tab:country_predictors}
\end{table*} 
\noindent\textbf{Occupations with opposite biases}: To explore the differences among the clusters, we define the concept of occupations with opposite biases. An occupation is considered opposite in two clusters if it is male-biased in one cluster and female-biased in the other or vice versa. To determine whether an occupation $o$ is opposite in two clusters $c_{1}$ and $c_{2}$, we consider four conditions:
\begin{itemize}
    \item occupation $o$ satisfies the condition in eq. (1) in cluster $c_{1}$ and the condition in eq. (3) in cluster $c_{2}$.
    \item occupation $o$ satisfies the condition in eq. (3) in cluster $c_{1}$ and the condition in eq. (1) in cluster $c_{2}$.
    \item occupation $o$ satisfies the condition in eq. (2) in $c_{1}$ and the condition in eq. (4) in cluster $c_{2}$. 
    \item occupation $o$ satisfies the condition in eq. (4) in cluster $c_{1}$ and the condition in eq. (2) in cluster $c_{2}$.
\end{itemize}
\noindent Table~\ref{tab:dissimilarities_clusters} displays the occupations following any of the four conditions mentioned above for the Global North and Global South cluster pairs. A careful inspection shows that the country-level indicators of GN-1 and GN-2, such as GDP per capita, gender gap, etc., as detailed in Table~\ref{tab:country_predictors} are very close and far apart from the GS cluster. Thus, we conclude that GN-1 and GN-2 are finer distinctions of a giant Global North cluster and, hence, are combined into GN for further inspection of occupations with opposite biases. An interesting observation from the table is that male-biased occupations in the Global North and female-biased occupations in the Global South seem to be intellectually driven, such as businessperson, diplomat, artist, photographer, songwriter, etc. This possibly suggests that women tend to do jobs that are perceived as ``soft'' in the Global South, and men are interested in such white-collar jobs in the Global North. In contrast, female-biased occupations in the Global North and male-biased occupations in the Global South seem to be more physical activity-based, e.g., basketball players, swimmers, field hockey players, choreographers, volleyball players, athletics competitors, etc. This possibly means that women are involved in physical jobs in the Global North, while men are interested in such jobs in the Global South. This surprising finding is indeed supported by recent publications. In the article on `Women's Employment' published in the journal \textit{Our World in Data}, the authors studied the employment of women worldwide in the labor market~\cite{ortiz2024women}. They observed that the female labor force participation is highest in the richest countries (primarily in the Global North) and lowest in medium-income countries (mostly Global South). Further authors reported that in the Global North, higher levels of economic development, industrialization, and gender equality have created greater opportunities for women to engage in various sectors, including physically demanding jobs\footnote{\url{https://tinyurl.com/3m8vuesx}}. Conversely, women in the Global South frequently encounter more significant barriers to labor force participation due to conservative social norms~\cite{klasen2019explains}, but they are notably present in informal sectors and less physically demanding roles. For instance, India has the highest percentage of female coding developers globally, with women comprising 22.9\% of the workforce~\footnote{\url{https://tinyurl.com/2tpm8enm}}.

\begin{table*} [ht]
    %\centering
    %\scriptsize
    \begin{tabular}{|p{3cm}|p{3cm}|p{3cm}|p{3cm}|p{3cm}|}
    \hline
        \textbf{Categories} & \textbf{TransE} & \textbf{DistMult} & \textbf{CompGCN} & \textbf{GeKC}\\
        \hline
        $(TPR_{m}>TPR_{f})_{GN}$, $(TPR_{m}<TPR_{f})_{GS}$ & athlete, businessperson, artist, radio personality, illustrator, translator, photographer, stage actor, economist, actor, songwriter, diplomat, biologist, physician, teacher & businessperson, physician, association football player, activist, film director, radio personality, teacher, water polo player, singer-songwriter, diplomat, television producer, judoka & swimmer, tennis player, zoologist, civil servant, dancer, artist, architect, amateur wrestler, visual artist, badminton player, athletics competitor, boxer, politician, radio personality, volleyball player, rugby union player, canoeist, academic, economist, farmer, field hockey player, pianist & academic, swimmer, television presenter, athlete, long-distance runner, judge, librarian 
 \\
         \hline
      $(TPR_{m}<TPR_{f})_{GN}$, $(TPR_{m}>TPR_{f})_{GS}$ & lawyer, scientist, autobiographer, television presenter, designer, physician, businessperson, athletics competitor, academic, film producer, entrepreneur, fashion designer, children's writer, photographer, field hockey player, sociologist, volleyball player, swimmer, choreographer, university teacher, basketball player  & musician, tennis player, zoologist, architect, university teacher, economist, politician, athletics competitor,  author, autobiographer, academic & diplomat, model, tennis player, photographer, zoologist, basketball player, amateur wrestler, autobiographer, poet, volleyball player, businessperson, botanist, judoka, engineer, biologist, physician & curator, rower, physician, official
 \\
         \hline
     $(FPR_{m}>FPR_{f})_{GN}$, $(FPR_{m}<FPR_{f})_{GS}$ & athletics competitor, lyricist, tennis player & businessperson, television actor, sculptor, singer-songwriter, television presenter & swimmer & - \\
         \hline
     $(FPR_{m}<FPR_{f})_{GN}$, $(FPR_{m}>FPR_{f})_{GS}$ & singer-songwriter, film producer, & swimmer, badminton player & sculptor, stage actor, taekwondo athlete & speed skater, teacher\\
         \hline
    \end{tabular}
    \caption{
    List of occupations that belong to opposite categories in the cluster pairs-- Global North (i.e., GN-1 and GN-2 together) and Global South for the attribute gender. By opposite category, we want to point out the occupations that are marked as male-biased in one cluster and female-biased in the other cluster of the cluster pairs. The pair of tuples under ``Categories'' in each row can be read as -- the first tuple is considered for the first cluster, i.e., Global North (GN), and the second one for the other cluster, i.e., Global South (GS). For example, the first row lists the occupations that belong to the fairness category $TPR_m > TPR_f$ in GN and $TPR_m < TPR_f$ in GS.} 
    %The second row denotes the occupations that comes under $TPR_m < TPR_f$ in the first cluster and $TPR_m > TPR_f$ in the second cluster. The third row indicates occupations with $TPR_m = TPR_f$ in both clusters but $FPR_m > FPR_f$ in the first cluster and $FPR_m < FPR_f$ in the second cluster. The last row contains occupations which signify that occupations have $TPR_m = TPR_f$ in both clusters but $FPR_m < FPR_f$ in the first cluster and $FPR_m > FPR_f$ in the second cluster. \pd{rephrase the caption.}}
    \label{tab:dissimilarities_clusters}
\end{table*}

%Our analysis resulted in the identification of two major clusters, which demonstrate a strong correlation between the classification of gender-biased and gender-neutral occupations and the economic and cultural differences between the two hemispheres. While we did not investigate the underlying reasons for these correlations, our findings suggest a potential relationship between societal biases that is hidden in the gender inclination of the occupations and the socio-economic structure of the identified clusters. Additionally,  These attributes reveal significant differences in socio-economic and cultural characteristics between the clusters, particularly between GN-1 and GS, as shown in Table~\ref{tab:country_predictors}.

%geographies in the first and second clusters showed greater progress in areas such as wealth, economic development, income inequality, democracy, and political stability compared to those in the third cluster, which we call Global South. Argentina is typically categorized as a country in the Global South, but our clustering approach grouped it into the Global North cluster. However, in this particular case, we can assume it as an outlier because it does not align with the commonly accepted classification. Interestingly, these clusters demonstrate the social biases that are hidden in the distribution of gender-dependent and gender-neutral occupations in different geographies and our pipeline is able to correctly identify them. 

\section{Conclusion}
In our study, we examine how biases embedded in a knowledge graph, specifically Wikidata, can influence the task of link prediction, which is essential for completing a knowledge graph. Specifically, we show that the presence of social bias in terms of protected attributes, for example-- gender and age in the graph, leads to certain occupations being classified as gender-inclined/age-inclined, even though they should be gender-neutral/age-neutral in an unbiased setting. We introduce a new dataset specifically designed for measuring social biases in knowledge graph completion tasks, sourced from Wikidata triples, incorporating two important sensitive attributes—gender and age. The dataset also includes a variety of occupations associated with human entities from 21 different regions globally. We quantify the biases in terms of the standard fairness metrics through our proposed noble framework-- AuditLP. An interesting finding of our study is that the extent of biases in occupations is related to the socio-economic division of the world, which separates developing and developed countries into two major groups, also known as the Global South and the Global North, respectively. Remarkably, the embedding learning algorithms, implemented from different genres -- TransE, DistMult, CompGCN and GeKC -- using which the AuditLP framework draws its features from yielded the same general results, despite their differences. Our conclusions are based on the optimal hyperparameter settings for these methods, with minor adjustments to these parameters not affecting the overall observations. So far, this work anchors on two sensitive attributes-- age and gender and 21 different geographical regions worldwide. As a future direction, we plan to implement our framework on other 
sensitive attributes, such as ethnicity, religion, etc., though extracting these attributes from Wikidata triples. Furthermore, the geographic scope of our dataset can be further expanded
to include additional regions. Upon acceptance, we shall release the full dataset and codes to facilitate future research.
%This work anchors on a single sensitive attribute -- gender. As a future direction, we plan to expand our investigation to examine the fairness implications of other sensitive attributes, such as race and religion.

\bibliographystyle{ACM-Reference-Format}
\bibliography{ref}

\iffalse
\section{Appendices}

If your work needs an appendix, add it before the
``\verb|\end{document}|'' command at the conclusion of your source
document.

Start the appendix with the ``\verb|appendix|'' command:
\begin{verbatim}
  \appendix
\end{verbatim}
and note that in the appendix, sections are lettered, not
numbered. This document has two appendices, demonstrating the section
and subsection identification method.

\fi
\end{document}